\documentclass[aps,prd, nofootinbib]{revtex4}

\usepackage{graphicx}
\usepackage{dcolumn}
\usepackage{amssymb}
\usepackage[all]{xy}
\usepackage{epsfig}
\usepackage{amsfonts,amsmath}
\usepackage{subfigure}
\usepackage{hyperref}

\newcommand{\half}{\frac{1}{2}}

\newcommand\beq{\begin{equation}}
\newcommand\eeq{\end{equation}}
\newcommand\bea{\begin{eqnarray}}
\newcommand\eea{\end{eqnarray}}
\newcommand\be{\begin{equation}}
\newcommand\ee{\end{equation}}
\newcommand\del{\partial}

\newcommand\al{\alpha}

\newcommand\K{{\cal K}}
\newcommand\R{{\cal R}}
\newcommand{\gsim}{\raise.3ex\hbox{$>$\kern-.75em\lower1ex\hbox{$\sim$}}}
\newcommand{\lsim}{\raise.3ex\hbox{$<$\kern-.75em\lower1ex\hbox{$\sim$}}}

\def\d{{\rm d}}
\newcommand\deriv[3][\d]{\frac{#1 #2}{#1 #3}}
\newcommand\partd[2]{\deriv[\del]{#1}{#2}}

\newcommand{\vfour}{\ensuremath{{\cal V}_4}}

\begin{document}

\title{Lessons from the decoupling limit of Ho\v rava gravity}

\author{ Ian Kimpton} 
\email[]{ppxik1@nottingham.ac.uk}
\author{Antonio Padilla} 
\email[]{antonio.padilla@nottingham.ac.uk}

\affiliation{School of Physics and Astronomy, 
University of Nottingham, Nottingham NG7 2RD, UK} 

\date{\today}

\begin{abstract}
We consider the so-called ``healthy" extension of  Ho\v rava gravity  in the limit where the Stuckelberg field decouples from the graviton.  We verify the alleged strong coupling problem in this limit, under the assumption that  no large dimensionless parameters are  put in by hand.  This follows from the fact that the dispersion relation for the Stuckelberg field does not have the desired  $z=3$ anisotropic scaling in the UV. To get the desired scaling and avoid strong coupling one has to introduce a low scale of Lorentz violation and retain some coupling between the graviton and the Stuckelberg field. We also make use of the foliation preserving symmetry  to show how the Stuckelberg field couples to some violation of energy conservation.  We source the Stuckelberg field using a point particle with a slowly varying mass and show that two such particles feel a constant attractive force. In this particular example, we see no Vainshtein effect, and violations of the Equivalence Principle. The latter is probably generic to other types of source and could potentially be used to place lower bounds on the scale of Lorentz violation.
\end{abstract}


\maketitle


%

%



%

%

\section{Introduction} \label{sec:intro}
Gravity is the most troublesome of fundamental forces. Although classical General Relativity describes gravity very well  over a large range of scales, one runs into real trouble at Planckian energies because the corresponding quantum theory is non-renormalisable. Interesting and {\it simple} proposals for a quantum theory of gravity are rare, which is why there has been so much interest in Ho\v rava's recent idea~\cite{horava1, horava2}.  

Roughly speaking, the non-renormalisability of GR can be associated with the fact that the propagator scales like $1/k^2$, and the coupling constant has negative mass dimension, $[G_N]=-2$. In contrast, in gravitational theories with higher derivatives,  the propagator falls off more quickly in the UV, and  the coupling constant can be non-negative~\cite{stelle}. As a result these theories are at least {\it power-counting} renormalisable, in contrast to General Relativity. Unfortunately, in relativistic theories, the higher derivatives typically introduce extra degrees of freedom and ghostly pathologies~\cite{higherdghosts, phantommenace}. In Ho\v rava's toy model,  this issue is avoided by breaking diffeomorphism invariance. This enables us to write down a theory that has only first order temporal derivatives, but contains higher order spatial derivatives.  The good UV behaviour is maintained, and there are no obvious  extra degrees of freedom and no obvious ghosts. Furthermore, relevant operators can be added to the Lagrangian so that the theory starts to look like General Relativity in the infra-red. Is this really a UV complete theory that approaches classical GR at low energies? Unfortunately not.

The key issue is broken diffeomorphism invariance. Diffeomorphism invariance is the {\it dynamical} symmetry of General Relativity, and not just a symmetry of the background. As such, it controls the number of propagating degrees of freedom that are present in the theory. It is well known that GR contains two spin 2 degrees of freedom. As diffeomorphism invariance is broken in Ho\v rava gravity, it is clear that it should contain extra degrees of freedom. The question is: what happens to these extra degrees of freedom as we approach the so-called ``GR limit"? It turns out that they typically become strongly coupled~\cite{us, blas1}. This casts serious doubts on  the UV completeness and therefore the success of the theory.

One might  hope to extend the theory in order avoid these problems, as they claim to do in the so-called ``healthy" extension of Ho\v rava gravity~\cite{blas2}. However, there has been some debate as to whether these healthy extensions are really that healthy. Sotiriou and Papazoglou considered the low energy limit of this theory and showed that it too suffered from problems with strong coupling~\cite{sotpap}. In response, the original authors have argued that the strong coupling scale might exceed the cut-off scale for the derivative expansions and be rendered meaningless~\cite{blas3}.  Obviously, one could force this to be the case by introducing a new scale in the However it is worth asking whether or not this is actually {\it necessary}, and if so, to what degree -- can the strong coupling scale be raised, or removed altogether,  simply by including higher order interactions, but without introducing large dimensionless  parameters?

Strong coupling seems to be an endemic feature of modified gravity models~\cite{dvalithm} and Ho\v rava gravity would appear to be no exception. However, phenomenologically, one might wish to take a more favourable view of strong coupling, especially if it can be linked to some sort of Vainshtein mechanism \cite{vain}. In DGP gravity~\cite{dgp, dgp+1, dgp+2, dgp+3}, for example, there is a longitudinal scalar mode that becomes strongly coupled at a scale of around $1000$ km \cite{dgpsc}. Far from being a disaster this strong coupling effect ensures that non-linear interactions become important sooner than expected, helping to screen the longitudinal scalar in the solar system so that one recovers General Relativity \cite{dgpvain}. It is only beyond the so-called Vainshtein scale, beyond the edge of the solar system where gravity is not so well tested, that the scalar is able to mediate an additional force.  Is there a Vainshtein mechanism at work in Ho\v rava gravity? Does it help with phenomenology?

In this paper we address these  and other issues directly by isolating the troublesome extra degree of freedom in each case, and studying its properties. This can be done using the ``Stuckelberg" trick to fully restore diffeomorphism invariance  to all orders, along the lines proposed in ~\cite{germani, blas1}. We will do this for all versions of the {\it non-projectable}\footnote{Non-projectable theories are those for which the lapse function depends on space~\cite{horava1, horava2}. We do not consider projectable theories for which the lapse is homogeneous, as these have been ruled out~\cite{blas1, bogdanos, KKarroja}. However, see \cite{projok} for a recent development.} theory, including  their ``healthy" extensions~\cite{blas2}. To drastically simplify the analysis we will take a limit in which the Stuckelberg field decouples completely from the spin 2 sector. The decoupled Stuckelberg theory ought to capture most  of the interesting physics associated with strong coupling, as well as the possible presence of ghosts and other pathologies\footnote{Surprisingly, it turns out one cannot hope to  capture the proposed resolution of strong coupling presented in \cite{blas3} unless one retains some coupling between the graviton and the Stuckelberg mode.}. Indeed we will show how many results can be recovered very simply using this decoupled limit of Ho\v rava gravity. In particular, we will show, very easily, that strong coupling is a problem even  for the so-called ``healthy" extensions, at least when one assumes that Lorentz violation occurs at the Planck scale. In other words, to have any hope of curing strong coupling one has to introduce a new scale by hand, below the would be strong coupling scale, pushing the scale of Lorentz violation to well below the Planck scale~\cite{blas3}.  We will also be able to study the role of non-linear interactions on the extra force mediated by the Stuckelberg field. For the case of a  point source with a slowly varying mass\footnote{Time variation in the mass is required to source the Stuckelberg field.}, we will find that there are possible violations of the Equivalence Principle, but no Vainshtein effect. 

The rest of this paper is organised as follows.  In the next section we write down the most general form for the action in the non-projectable theory, consistent with the reduced set of diffeomorphisms. We will restore full diffeomorphism invariance by performing a Stuckelberg trick along the lines proposed in ~\cite{germani, blas1}.  As an added bonus we shall use symmetry arguments to discover  how the Stuckelberg field couples to matter, in general. As suggested by the linearised analysis in \cite{us}, the Stuckelberg field couples to $\nabla_\mu T^{\mu\nu}$, which need not be zero in Ho\v rava gravity. In section \ref{sec:dec}, we take the decoupling limit, in which the Stuckelberg field and the spin 2 mode no longer interact.  This corresponds to taking the Planck length to zero,  but holding other length scales fixed, so that we are left with a non-linear theory of the Stuckelberg field propagating on Minkowski space. This simplification enables us to study important aspects of the theory in more detail in sections \ref{sec:fluc} and \ref{sec:vain}. In section \ref{sec:fluc} we will show that the so-called ``healthy" extension of Ho\v rava gravity suffers from a strong coupling problem, as suggested by \cite{sotpap}, at least if one assumes that no large dimensionless parameters are introduced by hand.  In some sense, our result is more robust than \cite{sotpap} since we also show that the scale of strong coupling lies below the cut-off scale in the derivative expansion.  Therefore to remove strong coupling we do indeed need to   introduce a new low energy scale in the higher derivative terms by brute force, lying well below the strong coupling scale calculated in the derivative expansion, as proposed by \cite{blas3}.  This introduces a scale of Lorentz violation that is well below the Planck scale.  In section \ref{sec:vain}, we consider some phenomenological aspects of Ho\v rava gravity.  We will focus on the force between two point particles with slowly varying masses. Owing, no doubt, to the level of symmetry, we will find no evidence for a Vainshtein effect, but we will see violations of the Equivalence Principle.  Finally, in section \ref{sec:conc}, we close with some concluding remarks. In particular we point out that experimental bounds on the E\"otv\"os parameter $\eta=2\frac{|a_1-a_2|}{|a_1+a_2|}$ could potentially be used to place lower bounds on the scale of Lorentz violation.

\section{Stuckelberging the general theory} \label{sec:gen}
In Ho\v rava gravity we break full diffeomorphism invariance by choosing a specific spacetime foliation along slices of constant time, $t$. We therefore make an ADM split, writing the full spacetime metric $g_{\mu\nu}$ in terms of the lapse $N(x, t)$, shift $N^i(x, t)$ and spatial metric $\gamma_{ij}(x, t)$,
\be 
ds^2=g_{\mu\nu}dx^\mu dx^\nu=-N^2 c^2 dt^2+\gamma_{ij}(dx^i+N^idt)(dx^j+N^jdt)
\ee
Note that by allowing the lapse to depend on space we are assuming that the theory is  {\it non-projectable}. Although full diffeomorphism invariance is broken, we retain the subgroup of diffeomorphisms that preserve the spacetime foliation
\be
t\to \tilde t(t), \qquad x^i \to \tilde x^i(x, t)
\ee
Let us now  consider a gravity action that is invariant under this subgroup. Since we do not want to introduce any pathological ghostly degrees of freedom, we require that each term in the kinetic part of the action  contains at most two time derivatives.  In contrast, in the potential part of the action,  one may have more than two spatial derivatives since full diffeomorphism invariance has been broken. If we have terms with up to $2z$ spatial derivatives, then time and space scale anisotropically in the UV, such that the scaling dimensions are given by 
\be
[x]=-1, \qquad [t]=-z
\ee
For a power counting renormalisable theory we need $z \geq 3$, which means  we must have  terms containing  at least six spatial derivatives~\cite{horava1, horava2}.   Here we consider the most general action for $z=3$,
\be
S=S_{grav}+S_m
\ee
where the gravitational part is given by, 
\be \label{horact}
S_{grav}=\frac{1}{\kappa^2} \int dt d^3x \sqrt{\gamma}N\left(K_{ij}K^{ij}-\lambda K^2\right)+S_{V}
\ee
and $S_m$ is the generalised matter action~\cite{horava1, horava2, blas2, sotvis}. Note that the extrinsic curvature of the spatial slices is given by
\be
K_{ij}=\frac{1}{2N}\left(\dot \gamma_{ij}-2D_{(i} N_{j)}\right)
 \ee
where dot is $\del/\del t$ and $D_i$ is the covariant derivative along the spatial slices. The potential part of the action, $S_V$, contains  contributions with six spatial derivatives that dominate in the UV, as well as relevant deformations that contain two and four spatial derivatives. Odd numbers of spatial derivatives are forbidden by spatial parity~\cite{sotvis}. The most general form for the potential, compatible with the reduced diffeomorphisms is given by
\be
S_{V}=\int dt d^3x \sqrt{\gamma}N c\left(\frac{V_2}{l_{UV}^2}+V_4+l_{UV}^2 V_6\right)
\ee
where $l_{UV}$ is some short distance Lorentz symmetry breaking scale,  and $c$ is the emergent speed of light. The two and four derivative terms are respectively given by
\be 
V_2=\alpha (a^i a_i)+\beta R
\ee
and
\be
 V_4=A_1 (a^i a_i)^2+A_2 (a^i a_i)a^j{}_j+A_3 (a^i{}_i)^2+A_4 a^{ij} a_{ij}  
  +B_1R a^i a_i+B_2R_{ij}a^ia^j+B_3R a^i{}_i+C_1R^2+C_2R_{ij}R^{ij}
\ee
where we introduce the notation $$a^{i_1 \ldots i_n}=D^{i_1} \cdots D^{i_n} \log N$$
Typically we assume that all the dimensionless coefficients appearing in the potentials are  $\lesssim {\cal O}(1)$. This seems reasonable on naturalness grounds by not introducing additional scales into the theory, other than the Lorentz symmetry breaking scale, which is expected to be Planckian\footnote{To resolve the issue of strong coupling along the lines suggested by \cite{blas3} one must introduce a low Lorentz violating scale, $M_* \ll M_{pl}$.}.   We make no attempt to write down the six derivative term, $V_6$, as it contains in excess of 60 different contributions. In any case, if we assume  that all the coefficients are $\lesssim {\cal O}(1)$ it turns out that  $V_6$ will not play a role in the decoupling limit. Note that $\beta$ is not physically relevant. Indeed,  if $\beta \neq 0$  we can set $\beta$  to be whatever we want simply by rescaling time. 

Let us now artificially restore full diffeomorphism invariance by performing a Stuckelberg trick~\cite{germani, blas1}. The first, albeit trivial, step is to undo the anisotropic scaling in the coordinates by introducing $x^0=ct$. This enables us to  formally refer to spacetime coordinates $x^\mu=(x^0, x^i)$, with each component having the same scaling dimension. Of course, this is not enough, because the spacetime foliations are still defined by $t=constant$ surfaces, explicitly breaking full diffeomorphism invariance. The trick is to redefine the foliation like so
$$
t=constant \qquad \to \qquad \phi(x)=constant
$$
 where $\phi$ is some arbitrary scalar function of the spacetime coordinates. The action can now be expressed in terms of covariant  tensors in four spacetime dimensions, as we will now explain.

We begin by defining the unit normal to the foliation,
$$\hat u^\mu= \frac{\nabla^\mu \phi}{\hat X}, \quad \hat X = \sqrt{-\nabla_\mu \phi \nabla^\mu\phi}$$
where $\nabla_\mu$ is the covariant derivative on the full spacetime. The induced metric on the spatial slices can now be expressed as a covariant $4D$ tensor, 
$$\gamma_{ij} \to \hat \gamma_{\mu\nu}= g_{\mu\nu}+\hat  u_\mu \hat u_\nu$$
This  takes the form of a  $4D$ projector on to a  spacelike submanifold, with unit normal $\hat u^\mu$. The extrinsic curvature can also be promoted to a covariant $4D$ tensor, since it is simply the Lie derivative of the induced metric along the direction of the normal. Specifically
$$\frac{1}{c}K_{ij} \to \hat \K_{\mu\nu}= \half {\cal L}_{\hat u} \hat \gamma_{\mu\nu}=\hat \gamma_{(\mu}^\alpha \hat  \gamma_{\nu)}^\beta  \nabla_\alpha \hat u_\beta$$
Without going into too much detail (see \cite{germani, blas1}), note that all the other terms can be promoted to  covariant $4D$ tensors in a similar way,
\bea
R_{ij} &\to& \hat \R_{\mu\nu}=\hat \gamma^\alpha_\mu \hat \gamma^\beta_\nu \gamma^{\rho \sigma}R^{(4)}_{\rho \alpha \sigma\beta}+\hat \K_{\mu \alpha}\hat  \K^\al_\nu-\hat \K \hat \K_{\mu\nu}\\
R &\to& \R=\hat \gamma^{\alpha\beta}\hat \gamma^{\rho \sigma}R^{(4)}_{\rho \alpha \sigma\beta}+\hat \K_{\mu \nu}\hat  \K^{\mu\nu}-\hat \K^2  \label{gcR1} \\
a^i &\to&\hat  a^\mu= \hat u^\beta \nabla_\beta \hat u^\mu \\
a^{ij} &\to & \hat a^{\mu\nu}=\hat \gamma^{(\mu}_\alpha  \hat \gamma^{\nu)}_\beta  \nabla^\alpha \hat a^\beta
\eea
We are now ready to write the Ho\v rava action in a completely covariant way. We find that
\be
S_{grav}=S_{GR}+\Delta S_K+\Delta S_2+\Delta S_4 +\Delta S_6+ S_{m}
\ee
where
$$S_{GR}=\frac{1}{l_{pl}^2}\int d^4 x \sqrt{-g} R^{(4)}$$
is the emergent GR piece, with corresponding Planck length $l_{pl}=\frac{~\kappa}{\sqrt{c}}$
$$\Delta S_K=\frac{1-\lambda}{l_{pl}^2}\int d^4 x \sqrt{-g} ~\hat \K^2$$
$$\Delta S_2=\frac{1}{l_{UV}^2}\int d^4 x \sqrt{-g} ~\alpha \hat a^\mu \hat a_\mu+\left(\beta- \frac{l_{UV}^2}{l_{pl}^2}\right)\hat \R$$ 
$$\Delta S_4=\int d^4 x \sqrt{-g} ~\hat {\cal V}_4$$
$$\Delta S_6=l_{UV}^2 \int d^4 x \sqrt{-g} ~\hat {\cal V}_6$$
where 
\be
\hat  {\cal V}_4=A_1 (\hat a^\mu \hat a_\mu)^2+A_2 (\hat a^\mu \hat a_\mu)\hat a^\nu{}_\nu+A_3 (\hat a^\mu{}_\mu)^2+A_4 \hat a^{\mu\nu} \hat a_{\mu
 \nu}  
  +B_1\hat \R \hat a^\mu \hat a_\mu+B_2\hat \R_{\mu\nu}\hat a^\mu \hat a^\nu+B_3\hat \R \hat a^\mu{}_\mu+C_1\hat \R^2+C_2\hat \R_{\mu\nu}\hat \R^{\mu\nu}
\ee
As before, we will not attempt to write down $\hat {\cal V}_6$.  At this stage we use the fact that the precise value of $\beta$ is physically irrelevant and set $\beta=  \frac{l_{UV}^2}{l_{pl}^2}$, thereby eliminating the last term in $\Delta S_2$.

The matter part of the action $S_{m}=S_m[g_{\mu\nu}, \phi, \Psi]$, where $\Psi$ denotes the contribution of the  matter fields. In principle matter couples to both the spacetime metric and the Stuckelberg field. It couples to the metric via the energy-momentum tensor, $T^{\mu\nu}=\frac{2}{\sqrt {-g}} \frac{\delta S_m}{\delta g_{\mu\nu}}$. Now, by requiring that the matter Lagrangian is invariant under foliation preserving diffeomorphisms, we find that 
\be
\hat \gamma_{\alpha \nu} \nabla_{\mu} T^{\mu\nu}   =0 \label{cons1}
\ee
and
\be
\frac{1}{\sqrt{- g}} \frac{\delta S_m}{\delta \phi}=-\frac{\hat u_\nu\nabla_\mu T^{\mu\nu}}{\hat X} \label{dsdp}
\ee
 As pointed out in \cite{us}, in the absence of full diffeomorphism invariance, we do not necessarily have  conservation of energy momentum, so the right hand side of equation (\ref{dsdp}) can be non-zero.  This corresponds to the  matter source in the Stuckelberg equation of motion. 

To see where these equations come from, let us first consider infinitesimal transformations of the form $x^\mu \to \tilde x^\mu=x^\mu+\xi^\mu$, with
\be
g_{\mu\nu}(x) \to \tilde  g_{\mu\nu}(\tilde x)=g_{\mu\nu}(\tilde x)-2\nabla_{(\mu} \xi_{\nu)}, \qquad \phi(x) \to \tilde \phi(\tilde x)=\phi(\tilde x)-\xi^\mu \del_\mu\phi \label{foltrans}
\ee
The matter action transforms as
\be
S_m \to S_m-\int d^4 \tilde x ~\left[2\nabla_{(\mu} \xi_{\nu)}\frac{\delta S_m}{\delta g_{\mu\nu}}+\xi^\mu \del_\mu\phi\frac{\delta S_m}{\delta \phi}\right]
\ee
The formulae (\ref{cons1}) and (\ref{dsdp}) follow from requiring that the matter action is  invariant under this transformation, for arbitrary $\xi^\mu$.  The important thing to note is that the transformation (\ref{foltrans}) is a foliation preserving transformation, since it preserves the relation $\phi(x)=const$.

\section{The decoupling limit} \label{sec:dec}
At large distances we expect  that gravity is mediated by a massless graviton propagating at the  emergent speed of light $c$ seen by all other particle species.   This is consistent with General Relativity and requires that $\lambda \to 1$ and $\alpha \to 0$ in the far infra-red.  However, in Ho\v rava gravity, the breaking of diffeomorphism invariance allows an additional scalar field, identified with the Stuckelberg field, $\phi$,  to take part in this interaction.   Furthermore, recent studies have suggested that this field becomes strongly coupled at a scale $\lesssim l_{pl}/\sqrt{|1-\lambda|}$~\cite{us, blas1}. Clearly it is  important to examine whether or not the additional scalar  leads to unacceptable phenomenology. 

Such a scenario is reminiscent from DGP gravity~\cite{dgp, dgp+1, dgp+2, dgp+3}.  In that case there is a  strongly coupled scalar and the  key physics can be seen by taking the decoupling limit in which the Planck length is sent to zero, but the scale of strong coupling remains fixed~\cite{dgpsc}. We will now do the same for Ho\v rava gravity. 
We begin by assuming that the Lorentz symmetry breaking scale is roughly Planckian, as one might expect on grounds of naturalness. Without any further loss of generality, we will take $l_{UV} = l_{pl}$ for the remainder of this paper. We now introduce two new scales
\be
l_\lambda= \frac{l_{pl}}{\sqrt{|1-\lambda|}}, \qquad l_\al= \frac{l_{pl}}{\sqrt{|\al |}}  \label{scales}
\ee
In the decoupling limit, $l_{pl} \to 0$, these scales will be held fixed.  Of course, this requires that $\lambda \to 1$ and $\alpha \to 0$ which is consistent with their expected running in the infra-red and  ensures that only $S_{GR}$ contributes to the graviton dynamics. However, by holding the scales (\ref{scales}) fixed we enable $\Delta S_K$ and $\Delta S_2$ to contribute to the Stuckelberg dynamics, along with $\Delta S_4$. The two dynamical sectors are completely decoupled as  $l_{pl} \to 0$, with all the  interesting phenomenology appearing in the Stuckelberg sector.  Note that even though $\al \to 0$, by fixing the scale $l_\al$, we can still consistently study the extended Ho\v rava gravity models proposed in \cite{blas2}. Of course, we can always allow $l_\al$  to diverge, if we wish to recover the original theory. 

 Let us now work out the details. We expand  about  Minkowski, in terms of a canonically normalised graviton, $h_{\mu\nu}$, like so\be
g_{\mu\nu}=\eta_{\mu\nu}+l_{pl}h_{\mu\nu}
\ee
It follows that
\be
S_{GR}=-\int d^4x ~\frac{1}{2}h^{\mu\nu} {\cal E}h_{\mu\nu}+{\cal O}(l_{pl})
\ee
\be
\Delta S_{K}=\frac{sgn(1-\lambda)}{l_\lambda^2}\int d^4x ~\K^2+{\cal O}(l_{pl})
\ee
\be
\Delta S_{2}=\frac{sgn(\alpha)}{l_\alpha^2}\int d^4x ~ a^\mu  a_\mu+{\cal O}(l_{pl})
\ee
\be
\Delta S_{4}=\int d^4x ~ {\cal V}_4+{\cal O}(l_{pl})
\ee
\be
\Delta S_{6}={\cal O}(l_{pl}^2)
\ee
\be
S_m=S_m[\eta_{\mu\nu}, \phi, \Psi]+l_{pl}\int d^4x ~\frac{1}{2}h_{\mu\nu} T^{\mu\nu}+{\cal O}(l_{pl}^2)
\ee
where $ {\cal E}h_{\mu\nu}$ is the Einstein tensor linearised about Minkowski, and 
\be
 \K_{\mu\nu}=\left(\delta^\al_\mu+\frac{\del^\al \phi \del_\mu \phi}{ X^2}\right)\left(\delta^\beta_\nu+\frac{\del^\beta \phi \del_\nu \phi}{ X^2}\right)\frac{\del_\al \del_\beta \phi}{ X}, \qquad 
 \K=\eta^{\mu\nu}  \K_{\mu\nu} 
\ee
\be
 X=\sqrt{-\del_\mu\phi \del^\mu\phi},
\ee
\be
 a_\mu=\left(\delta^\al_\mu+\frac{\del^\al \phi \del_\mu \phi}{ X^2}\right)\frac{\del^\beta \phi \del_\al \del_\beta \phi}{ X^2}
\ee
\bea
  {\cal V}_4&=&A_1 ( a^\mu  a_\mu)^2+A_2 ( a^\mu  a_\mu) a^\nu{}_\nu+A_3 ( a^\mu{}_\mu)^2+A_4  a^{\mu\nu}  a_{\mu
 \nu}  \nonumber\\ &&+B_1(   \K_{\mu\nu} \K^{\mu\nu} - \K^2)  a^\mu  a_\mu+B_2( \K_{\mu \alpha}  \K^\al_\nu- \K  \K_{\mu\nu}) a^\mu  a^\nu \nonumber \\
&&  +B_3(  \K_{\mu\nu} \K^{\mu\nu} - \K^2) a^\mu{}_\mu +C_1( \K_{\mu\nu} \K^{\mu\nu} - \K^2)^2
\nonumber \\
&&
+C_2( \K_{\mu \alpha}  \K^\al_\nu- \K  \K_{\mu\nu})( \K^{\mu}_{\beta}  \K^{\beta \nu}- \K  \K^{\mu\nu})
\eea
\be
a_{\mu\nu}=\left(\delta^\al_\mu+\frac{\del^\al \phi \del_\mu \phi}{ X^2}\right)\left(\delta^\beta_\nu+\frac{\del^\beta \phi \del_\nu \phi}{ X^2}\right)\del_{(\al} a_{\beta)}
\ee
Variation of the matter action with respect to the Stuckelberg field gives
\be
\frac{\delta S_m}{\delta \phi}=-\frac{ u_\nu\del_\mu T^{\mu\nu}}{X} +{\cal O}(l_{pl})
\ee
where $u_\mu=\del_\mu \phi/ X$.  The violation of energy-momentum   conservation has some characteristic scale $\Gamma$, so that schematically $\nabla T^{\mu\nu} \sim \Gamma T^{\mu\nu}$. We will typically take $\Gamma$ to be much smaller than the overall scale of the energy-momentum tensor. Indeed, by taking $\Gamma \lesssim H_0$, where $H_0$ is the current Hubble scale we can argue that violations of energy-momentum conservation would not have been detected during the universe's lifetime. 

If we now take the decoupling limit,
\be
l_{pl} \to 0,\qquad \al \to 0, \qquad \lambda \to 1, \qquad  T^{\mu\nu} \to \infty, \qquad  \Gamma\to 0, \label{limit}
\ee
whilst keeping $l_\al$, $l_\lambda$, $l_{pl}T^{\mu\nu}$ and $\Gamma T^{\mu\nu}$ finite, 
we arrive at the decoupled Ho\v rava action
\be
S=S_{graviton}+S_{stuckelberg}
\ee
where
\be
S_{graviton}=-\int d^4x ~\frac{1}{2}h^{\mu\nu} {\cal E}h_{\mu\nu}+l_{pl}\int d^4x ~\frac{1}{2}h_{\mu\nu} T^{\mu\nu}
\ee
and
\be\label{stuckaction}
S_{stuckelberg}=\int d^4x \left[\frac{sgn(1-\lambda)}{l_\lambda^2} \K^2+\frac{sgn(\alpha)}{l_\alpha^2} a^\mu  a_\mu +  {\cal V}_4\right]+S_m[\eta_{\mu\nu}, \phi, \Psi]
\ee
The equations of motion for the Stuckelberg field can be derived from the action \eqref{stuckaction},
\beq \label{eomphi}
\del_\nu \left( \gamma^{\mu \nu} \frac{\rho_\mu}{X} \right) =\frac{ u_\nu\del_\mu T^{\mu\nu}}{X} ,
\eeq
where $\gamma^{\mu\nu}=\eta^{\mu\nu}+u^\mu u^\nu$ and 
\be
\rho^{\mu} = \lambda^{\nu} \del^{\mu} u_{\nu} - \del_{\nu} \left( u^{\nu} \lambda^{\mu} \right)  - \del_\nu \left( \mu^{\mu \nu} \right) + a_{\nu} \mu^{\mu \nu}  
 {}  + 2 \lambda^{\rho \sigma} u_{(\rho} \gamma_{\sigma ) \nu}  \del^\mu a^\nu + 2 \lambda^{\mu \sigma} u^{(\al} \gamma^{\beta )}_{\sigma}  \del_{\al} a_{\beta}   ,
\ee
It is clearer to leave $\rho^\mu$ written implicitly in terms of the following
\bea
\lambda^{\mu} &=& \del_{\nu} \left( \partd{\vfour}{a_{\rho \sigma}} \gamma^{\nu}_{( \rho} \gamma^{\mu}_{\sigma )} \right) - 2 \frac{sgn(1-\lambda)}{l_\lambda^2} \K u^{\mu} - \partd{\vfour}{\K_{\mu \nu}} u_{\nu}  - 2\frac{sgn(\alpha)}{l_\alpha^2} a^{\mu} - \partd{\vfour}{a_{\mu}} , \\
\lambda^{\mu \nu} &=& -\partd{\vfour}{a_{\mu \nu}}, \\
\mu^{\mu \nu} &=& - 2 \frac{sgn(1-\lambda)}{l_\lambda^2} \K \eta^{\mu \nu} - \partd{\vfour}{\K_{\mu \nu}} .
\eea
The derivatives of the potential are given in Appendix \ref{sec:potdevs}.  The right hand side of equation (\ref{eomphi}) goes like $\Gamma T^{\mu\nu}$ and, as such, remains finite.  Note further that the Stuckelberg equation of motion (\ref{eomphi}) is invariant under $\phi \to f(\phi)$, as required by foliation preserving diffeomorphisms. In principle one  could use this equation to study the response of the Stuckelberg field to the presence of a non-trivial source.  In particular one can ask whether or not the Stuckelberg field gets screened at short distances due to the Vainshtein effect. We will return to this issue in section \ref{sec:vain}.

\section{Fluctuations in the Stuckelberg field} \label{sec:fluc}
In this section we will consider fluctuations of the Stuckelberg field about the  vacuum, $\phi = \bar \phi + \chi$,  where $ \bar \phi = x^0$. This choice of vacuum  corresponds to choosing a constant time foliation of Minkowski space.  To study the dynamics of vacuum fluctuations we simply expand the action (\ref{stuckaction}) order by order in $\chi$, neglecting for now the contribution from the matter sector. 
 The result is 
\be
S_\chi=\sum_{n=2}^\infty S_n[\chi]
\ee
where $S_n[\chi]$ is of order $\chi^n$. On this background, $\bar a_\mu = \bar a_{\mu \nu} = \bar \K_{\mu \nu} = 0$. This drastically reduces the number of terms that appear in the expansion especially at low order. At quadratic and cubic order we find, 
\beq \label{ugquadaction}
S _2 [\chi]= \int \d^4 x \left\{ \frac{sgn (1 - \lambda)}{l_\lambda^2} (\del^2 \chi)^2 + \frac{sgn (\al)}{l_\al^2} (\del_i \dot \chi)^2 + (A_3+ A_4) (\del^2 \dot \chi)^2 \right\}
\eeq
where dot is now $\del_0$,   the spatial Laplacian  is $\del^2 \equiv \del_i \del_i$, and
\bea
S_3 [\chi]&=& \int \d^4 x \left\{  2 \frac{sgn (1 - \lambda)}{l_\lambda^2} \left[ - 2 (\del_i \dot{\chi})(\del_i \chi)(\del^2 \chi) - \dot{\chi}(\del^2 \chi)^2 \right] \right. \nonumber \\
&&{} + 2 \frac{sgn (\al)}{l_\al^2} \left[ - \ddot{\chi} (\del_i \chi)(\del_i \dot{\chi}) - \dot{\chi}(\del_i \dot{\chi})^2 - (\del_i \dot{\chi})( \del_j \chi)(\del_i \del_j \chi ) \right] \nonumber\\
&&{} - A_2 (\del_i \dot{\chi})^2 \del^2 \dot \chi \nonumber\\
&&{} - 2 A_3 \left[ (\del_i \ddot{\chi})(\del_i \chi)(\del^2 \dot \chi) + \ddot \chi (\del^2 \chi)(\del^2 \dot \chi) + (\del_i \dot \chi)^2 \del^2 \dot \chi + \dot \chi (\del^2 \dot \chi)^2 \right. \nonumber \\
&&{} \qquad \left. + (\del^2 \dot \chi)(\del_i \del^2 \chi)(\del_i \chi) + (\del_i \del_j \chi)^2 \del^2 \dot \chi \right] \nonumber \\
&&{} - 2 A_4 \left[ (\del_i \ddot \chi)(\del_j \chi)(\del_i \del_j \dot \chi) + \ddot \chi (\del_i \del_j  \chi) (\del_i \del_j \dot \chi) + (\del_i \dot \chi)(\del_j \dot \chi)(\del_i \del_j \dot \chi) \right. \nonumber \\
&&{} \qquad \left. + \dot \chi (\del_i \del_j \dot \chi)^2 + (\del_i \del_j \dot \chi) (\del_i \del_j \del_k \chi) \del_k \chi + (\del_i \del_k \chi)(\del_j \del_k \chi)(\del_i \del_j \dot \chi) \right] \nonumber \\
&&{} \left. -B_3 \left[ (\del_i \del_j \chi)^2 \del^2 \dot \chi - (\del^2 \chi)^2 \del^2 \dot \chi \right] \phantom{\frac{0}{0}} \right\} \label{ugcubaction}
\eea
The quartic term also contains the following contribution
\be
S_4[\chi] \supset  \int \d^4 x \frac{sgn (\al)}{l_\al^2} \left[\del_j \chi \del_i\del_j \chi \right]^2 \label{s4}
\ee
We are now in a position to determine the conditions on $\lambda$ and $\al$ to ensure the theory is free from any number of pathologies including ghosts (which violate unitarity), tachyons (which lead to instabilities), and superluminal mode propagation (which violates causality). In addition, we will check the scale of strong-coupling and compare it to the cut-off scale for the low energy effective theory. 
\subsection{Ghosts, tachyons and superluminal propagation} \label{sec:ghosts}
Let us begin by exorcising the ghost and throwing out the tachyon. To this end, it is convenient to rewrite the  quadratic action \eqref{ugquadaction} as follows
\beq \label{ugquadaction1}
S _2 [\chi]= \int \d^4 x \left\{ \dot \chi\left(\frac{sgn (\al)}{l_\al^2}\Delta^2+(A_3+ A_4) \Delta^4\right)\dot \chi + \frac{sgn (1 - \lambda)}{l_\lambda^2} \chi\Delta^4 \chi \right\}
\eeq
where we have introduced the operator $\Delta=\sqrt{-\del^2}$, which measures the magnitude of momentum. To avoid a ghost, we require the kinetic term in the action to be positive, and so
\be
\frac{sgn (\al)}{l_\al^2}+(A_3+ A_4) \Delta^2 >0
\ee
At low energies, this means that we require $\al>0$ to avoid the ghost, whereas at high energies we require $A_3+A_4>0$. It now follows that a tachyonic instability will kick in unless $sgn(1-\lambda)<0$, or in other words $\lambda>1$. So, in summary to avoid both ghosts and tachyons we require
\[
\al > 0 \qquad \lambda > 1 .
\]
This can be contrasted with the result obtained in \cite{blas2}, where the conditions were
\[
0 < \al < 2 \qquad \lambda > 1 \text{ or } \lambda < 1/3 .
\]
The difference arises due to taking the decoupling limit, in which $\al \to  0$, $\lambda \to 1$, although our results are clearly consistent.

Let us now consider the possibility of superluminal propagation\footnote{In a Lorentz violating theory, superluminal modes are not necessarily an issue. However, in the far infra-red, our effective theory is designed to be approximately Lorentz invariant, so it is desirable, if not essential, to prohibit superluminal propagation at low energies.}.  To get a handle on the speed at which the Stuckelberg mode propagates, we consider the linearised equation of motion 
\beq
\frac{sgn (1 - \lambda)}{l_\lambda^2} \del^2 (\del^2 \chi) + \frac{sgn (\al)}{l_\al^2} (\del^2 \ddot \chi) - (A_3 + A_4) \del^2 \del^2 \ddot \chi = 0.
\eeq
and derive the dispersion relation
\be
w^2=-\frac{sgn(1-\lambda)}{sgn(\al)}\left(\frac{l_\al}{l_\lambda}\right)^2\frac{ k^2}{1+(A_3+A_4)sgn(\al)l_\al^2 k^2} \label{disp}
\ee
At low energies $k <1/l_\alpha$, the wave propagates with sound speed given by
\beq
c_s^2  =-\frac{sgn(1-\lambda)}{sgn(\al)}\left(\frac{l_\al}{l_\lambda}\right)^2= \frac{\lambda-1}{\al} \label{cs}
\eeq
Note that the sound speed is real in the absence of ghosts and tachyons. In addition,  to avoid superluminal propagation we also require $c_s \leq 1$, which gives 
\be
l_\al \leq l_\lambda \qquad \implies \qquad \left| \lambda - 1 \right| \leq \left| \al \right|. 
\ee
Again, we can contrast our expression (\ref{cs}) with the sound speed given in \cite{blas2, sotpap},
\be
 c_s^2{}|_\text{exact}= \frac{2-\alpha}{\al} \left(\frac{\lambda-1}{3\lambda-1}\right)
\ee
This expression was derived by working directly with the original action (\ref{horact}), without any Stuckelberg tricks. However, once again we see that it is consistent with the expression (\ref{cs}) derived here in the decoupling limit $\lambda \to 1, ~\al \to 0$.  This demonstrates both the power and limitations of the Stuckelberg trick in  the decoupling limit. It is clear that working directly with Stuckelberg action (\ref{stuckaction}) reveals the key physics more easily than working with the full Ho\v rava action (\ref{horact}).  This truncation works perfectly well as long as we are happy to stay close to the limiting values of $\lambda$ and $\al$.

From now on we will assume that $\lambda>1$, $\al>0$ to avoid ghosts and tachyons, and  that the low energy speed of sound is given by $c_s \equiv l_\al / l_\lambda \leq 1$.

\subsection{Strong coupling}
Blas {\it et al} recently claimed that the so-called ``healthy" theory~\cite{blas2} might be free from strong-coupling~\cite{blas3}, contrary to the claims made in~\cite{sotpap}.  Their argument roughly goes as follows. They point out that the analysis of \cite{sotpap} only includes the lowest order terms in the derivative expansion, corresponding to two-dimensional operators.  Therefore, the effective theory is only valid at energies below some scale $\Lambda_{hd}$. If the strong coupling scale derived by \cite{sotpap} exceeds this cut off, then it cannot be taken seriously since higher order operators should have been included in the analysis~\cite{blas3}. They give a toy example in which an erroneous strong coupling scale is derived in the effective theory, only to disappear when the higher order operators are included.  

Of course, for the case of Ho\v rava gravity, one can force a suitably low cut-off in the derivative expansion by hand, by introducing a low energy scale, $M_*$, in the higher derivative terms at quadratic order. This scale corresponds to the new cut-off in the derivative expansion, and should lie below the would-be strong coupling scale. In some sense this is reminiscent of string theory in that the string scale is introduced just below the Planck scale where strong coupling would otherwise occur in gravity.   According to~\cite{sotpap}, the low energy effective theory becomes strongly coupled at a scale $ \frac{1}{l_\alpha} \sim \sqrt{\alpha} M_{pl}$, so this suggests we should  take $ M_* <\sqrt{\alpha} M_{pl}$.   As experimental constraints require $\alpha \lesssim 10^{-7}$, this forces $M_*$ to lie three to four orders of magnitude below the Planck scale~\cite{blas3}.  If the new scale appears at order $2z$ (in spatial derivatives), we must introduce a dimensionless parameter of order $B \sim  \alpha  (\frac{M_{pl}}{M_*})^{2z-2}>\alpha^{2-z}$.  The proposal to avoid strong coupling requires the scalar mode to enter a phase of  anisotropic scaling,  with dispersion relation $w^2 \propto k^6$,  before strong coupling kicks in the low energy effective theory~\cite{blas3}.    To get the right anisotropic scaling at the right scale, we therefore need to introduce a large term at $z=3$, with  dimensionless parameter $B >\frac{1}{\alpha} 
\gtrsim 10^{7}$. 

 In the context of this paper, this proposal corresponds to introducing a large dimensionless coefficient in the potential (specifically, the six derivative piece, ${\cal V}_6$). We did not consider this possibility when taking the decoupling limit, preferring instead to keep all dimensionless coefficients of order one, on grounds of naturalness. However, even if we had introduced some new scale $M_* \ll M_{pl}$ and taken the limit $l_{pl} \to 0$ whilst holding $M_*$ fixed, the dispersion relation for the Stuckelberg mode would not have coincided with the  desired anisotropic scaling for the scalar mode in \cite{blas3},  $w^2 \sim k^6/M_*^4$ for large $k$. To recover this behaviour using the  Stuckelberg approach, we need to retain some coupling between the graviton and the Stuckelberg mode, at least to quadratic order. This is beyond the scope of the current paper, but it does illustrate some of the  limitations of the decoupling limit. 

Nonetheless, it is still worth asking whether or not this brute force approach is absolutely necessary, and to what degree. Can the strong coupling scale be raised, or removed altogether,  simply by including higher order interactions, but without introducing large coefficients? Whilst this might be possible in principle\footnote{For example, consider the following toy model, in which we have small kinetic term at low energies, as in Ho\v rava gravity, but with no additional large parameters are introduced to the higher order Lorentz symmetry breaking terms,
$$
S=\int d^4 x~\frac{1}{2} \epsilon^2 M_{pl}^2 \left[ \dot \psi^2-\psi \Delta^2 \psi\right]+\psi \Delta^4 \psi + \frac{1}{M_{pl}^2}(\Delta^2 \psi)^3+ \frac{1}{M_{pl}^2}\psi(\Delta^2 \psi)^3, \qquad \epsilon \ll 1
$$
In the low energy theory, the relativistic kinetic term dominates, and the dominant interaction becomes strongly coupled at the scale  $\Lambda_\text{false} \sim \epsilon^{2/3} M_{pl}$, which is above the low energy cut-off at $\epsilon M_{pl}$.  By studying the theory at higher energies $\Delta > \epsilon M_{pl}$, it can be shown that the dominant interaction  actually becomes strongly coupled at the higher scale $\Lambda_\text{true} \sim \epsilon^{1/3} M_{pl} >\Lambda_\text{false}$.}
let us demonstrate explicitly how it is not the case here, at least at the level of the decoupled Stuckelberg theory. To this end, and as we have already emphasised, we will assume that  all  coefficients $A_i$, $B_i$ and $C_i$ are ${\cal O}(1)$, as are combinations of these coefficients.  This enables us to make definite statements in what follows, but is also to be expected on grounds of naturalness.  In contrast to \cite{sotpap}, however, we will not necessarily restrict attention to the low energy effective field theory.

Now, the first thing to do is to perform a derivative expansion at quadratic order in order to establish both the cut-off and the leading order terms in the effective theory.  We find that
\be
S_2[\chi]=\int \d^4 x \left\{\frac{1}{l_\al^2} \dot \chi\Delta^2\dot \chi - \frac{1}{l_\lambda^2} \chi\Delta^4 \chi \right\}\left(1+{\cal O}(\Delta^2l_\al^2)\right)
\ee
Clearly then the cut off for the effective theory is given by
\be
\Lambda_{hd} \sim \frac{1}{l_\al}
\ee
Cubic and quartic order interaction terms are given by equations (\ref{ugcubaction}) and (\ref{s4}).  Generically, there are three types of terms appearing at order $n$, corresponding to each of the first three terms in the Stuckelberg action (\ref{stuckaction}). Using dimensional analysis one can easily show that these terms are schematically given by
\bea
S^\al_{(n, a)}[\chi] &=& \int \d^4 x ~\frac{1}{l_\al^2} (\del_0)^a \Delta^{n+2-a} \chi^n \\
S^\lambda_{(n, a)}[\chi] &=& \int \d^4 x ~\frac{1}{l_\lambda^2} (\del_0)^a \Delta^{n+2-a} \chi^n \\
S^V_{(n, a)}[\chi] &=& \int \d^4 x  ~(\del_0)^a \Delta^{n+4-a} \chi^n
\eea
where $a$ controls the number of time derivatives. To estimate the scale at which these terms become strongly coupled, if at all, we first need to canonically normalise the quadratic part of the action. To this end we set
$$
\hat x^0=c_s x^0, \qquad \hat x_i=x_i, \qquad \hat \chi=\frac{\sqrt{c_s}}{l_\al}\chi
$$
so that the quadratic part of the action becomes
\be
S_2[\hat \chi]=\int \d^4  \hat x ~ \del_{\hat 0} \hat \chi\Delta^2\del_{\hat 0} \hat \chi - \hat \chi\Delta^4 \hat \chi 
\ee
It follows that the interaction terms now go like
\bea
S^\al_{(n, a)}[\hat \chi] &=& \int \d^4 x ~\left(\frac{1}{\Lambda^\al_{(a, n)}}\right)^{n-2} (\del_{\hat 0})^a \Delta^{n+2-a} \hat \chi^n \\
S^\lambda_{(n, a)}[\hat \chi] &=& \int \d^4 x~\left(\frac{1}{\Lambda^\lambda_{(a, n)}}\right)^{n-2} (\del_{\hat 0})^a \Delta^{n+2-a} \hat \chi^n \\
S^V_{(n, a)}[\hat \chi] &=& \int \d^4 x ~\left(\frac{1}{\Lambda^V_{(a, n)}}\right)^{n}  (\del_{\hat 0})^a \Delta^{n+4-a} \hat \chi^n \\
\eea
where
\be \label{strongcouplingscales}
\Lambda^\al_{(a, n)}=\frac{1}{l_\al}c_s^{\frac{1}{2}+\frac{2-a}{n-2}}, \qquad
\Lambda^\lambda_{(a, n)}=\frac{1}{l_\al}c_s^{\frac{1}{2}-\frac{a}{n-2}}, \qquad
\Lambda^V_{(a, n)}=\frac{1}{l_\al}c_s^{\frac{1}{2}+\frac{1-a}{n}}
\ee
These scales are the scales at which the corresponding terms become strongly coupled. However, it is important to note that not all of these terms actually appear in the action, as one can immediately see by looking at the cubic term (\ref{ugcubaction}). Given that $c_s \leq 1$, it turns out (see Appendix \ref{sec:lowestscale}) that the lowest of these scales to appear in the action is given by 
\be
\Lambda^\al_{(1,3)}= \Lambda^\al_{(0, 4)}=\frac{1}{l_\al}c_s^{3/2}
\ee
This is the scale at which the largest interaction terms become significant and one enters a strongly coupled regime. The strongly coupled terms correspond to the cubic interaction,
$$
-2 \int \d^4 x~ \frac{sgn (\al)}{l_\al^2} (\del_i \dot{\chi})( \del_j \chi)(\del_i \del_j \chi ) 
$$ 
and the quartic interaction given by equation (\ref{s4}).  We therefore identify the lowest energy strong coupling scale as being
\be
\Lambda_{sc} \sim \frac{1}{l_\al}c_s^{3/2}
\ee
Now, since $c_s \leq 1$, it follows that $\Lambda_{sc} \lesssim \Lambda_{hd} \sim 1/l_\al$, which means that the derived strong coupling scale does indeed lie below the cut off of the derivative expansion.  Of course, one may call into question this conclusion if $c_s \sim 1$, as then we have $\Lambda_{sc} \sim \Lambda_{hd}$. To allay any possible concerns let us consider what happens at high energies $\Delta \gg 1/l_\al$.  Then the quadratic part of the action is given by
\be
S _2 [\chi]= \int \d^4 x \left\{ \left(A_3+A_4\right) \dot \chi\Delta^4 \dot \chi - \frac{1}{l_\lambda^2} \chi\Delta^4 \chi \right\}\left(1+{\cal O}(1/\Delta^2l_\al^2 )\right)
\ee
As expected $x^0$ and $x^i$ scale differently in the UV,
$$
x^i \to b^{-1}x^i, \qquad x^0 \to x^0
$$
The fact that $x^0$ does not scale makes sense given that the dispersion relation (\ref{disp}) goes like $w \sim constant$ for large $k$. Indeed, to quadratic order the system reduces to a simple harmonic oscillator with fixed frequency of oscillation. 

In order to keep $S_2[\chi]$ invariant under the scaling, we must have
$$\chi \to b^{-1/2}\chi$$
It follows that the interaction terms scale like so
\bea
S^\al_{(n, a)}[\chi] &\to&b^{\frac{n}{2}-a-1}S^\al_{(n, a)}[\chi]  \\
S^\lambda_{(n, a)}[\chi] &\to&b^{\frac{n}{2}-a-1} S^\lambda_{(n, a)}[\chi]  \\
S^V_{(n, a)}[\chi] &\to&b^{\frac{n}{2}-a+1}S^V_{(n, a)}[\chi]  
\eea
These interactions become relevant in the UV whenever the exponent of $b$ is positive in the above scaling. Interactions with many time derivatives (that is, with large $a$),  are irrelevant in the UV, whereas those with fewer time derivatives become relevant. Indeed, the fourth order  interaction term given by equation (\ref{s4}) is clearly relevant, as it scales like $b^3$. Therefore there is no reason to expect that  the theory is  UV finite, even for $c_s \sim 1$. To make sense of the perturbative theory we need to introduce a cut-off given by the strong coupling scale. For $c_s \sim 1$, the only scale we have available is $1/l_\al \sim 1/l_\lambda$, so it follows that this corresponds to the scale at which equation (\ref{s4}) becomes large.

In conclusion then, unless we introduce some new scales by brute force, the ``healthy"  theory is unlikely to be  UV finite since it becomes strongly coupled at a scale 
\be
\Lambda_{sc} \sim \frac{1}{l_\al}c_s^{3/2}=\left(\frac{l_\al}{l_\lambda^3}\right)^{1/2}=\frac{1}{l_{pl}}\left[\frac{(\lambda-1)^3}{\al}\right]^{1/4} \label{scscale}
\ee
This result agrees with \cite{sotpap},  at least when $c_s \sim 1$, but is more robust, having considered the  effect of higher derivative corrections and allowing for $c_s \ll 1$. The correct interpretation of this result is to realise that we must introduce new physics by hand, below the scale $\Lambda_{sc}$. We can do this by explicitly introducing a new low scale of Lorentz violation, $M_*$,  in the higher derivative terms, as proposed in \cite{blas3}. Experimental considerations actually push $\Lambda_{sc}$, and by association $M_*$,  to well below the Planck scale.

 \section{The Stuckelberg force} \label{sec:vain}
It has been suggested that strong coupling problems in some versions of Ho\v rava gravity might be a blessing in disguise, at least from a phenomenological perspective. The claim is that a Vainshtein mechanism might occur, such that non-linear interactions become important, helping to screen any additional force due to the Stuckelberg mode. In this section we will derive the size of the Stuckelberg force, and compare it to the size of the usual force mediated by the graviton.

Of course, we need a suitable source. To excite the Stuckelberg mode, this must violate the usual energy-momentum conservation law, as is clear from equation (\ref{eomphi}). A simple choice is a time dependent point mass, with energy-momentum tensor
\be
T^{\mu\nu}=M(x^0) \delta^{(3)}(\vec x)\text{diag}(1, 0, 0,0)
\ee
Recall that the violation of energy-momentum conservation is characterised by some scale $\Gamma$, which we will take to be much less than overall scale of the energy-momentum tensor. In other words
$$\Gamma \sim \frac{|\del_0 M|}{M} \ll M$$
We will consider the following simple cases: a slowing decaying point mass,
\be
M(x^0)=M_* \exp(-\Gamma x^0)
\ee
or a slowly oscillating point mass,
\be
\ \qquad M(x^0)=M_* (1-\sin(\Gamma x^0))
\ee
In the decoupling limit given by equation (\ref{limit}), we see that we must take $M_* \to \infty$, $\Gamma \to 0$ and $l_{pl}\to 0$, holding $l_{pl}M_*$ and $\Gamma M_*$ fixed.  At this stage we could, in principle, solve the Stuckelberg equation of motion (\ref{eomphi}) to leading order, but we can do better than that. We can take advantage of the strong coupling discussed in the previous section to simplify the full non-linear analysis, as we will now describe.

Recall that  fluctuations on the trivial vacuum become strongly coupled at a scale $\Lambda_{sc}$,  given by equation (\ref{scscale}).  All the features of this strongly coupled theory can  be captured by taking the limit $l_\al \to 0$, $l_\lambda \to 0$, whilst holding $\Lambda_{sc}$ fixed.  This just means that $l_{pl} \to 0$ faster than $\lambda \to 1$ and $\al \to 0$, so in a sense it corresponds to  the case where deviations from General Relativity play a maximal role\footnote {In the opposite limit, $l_\lambda \to \infty$, $l_\al \to \infty$, there is no deviation from GR whatsoever at low energies, since we might as well just set $\lambda=1$, $\al=0$ from the outset.}. Note that this implies that  the  speed of sound $c_s \to 0$.  As regards the scaling of matter in this limit, we will assume that both $l_{pl}T^{\mu\nu}/\sqrt{c_s}$ and $\Gamma T^{\mu\nu}$ remain finite so that both the graviton and the Stuckelberg sector get non-vanishing source terms, as we will show presently.

In this limit all but the largest interaction terms discussed in the previous section go away, and the full theory is reduced to
\be
S_{\hat \chi}=\int \d^4  \hat x ~ \left[\del_{\hat 0} \hat \chi\Delta^2\del_{\hat 0} \hat \chi - \hat \chi\Delta^4 \hat \chi - \frac{2}{\Lambda_{sc}} \del_{\hat 0} \del_i \hat \chi \del_i \del_j \hat \chi \del_j \hat \chi +  \frac{1}{\Lambda_{sc}^{2}}  \left[\del_j \hat \chi \del_i\del_j \hat \chi \right]^2\right] -\int \d^4  \hat x ~ \frac{\hat \chi}{\Lambda_{sc}} \del_\mu T^{\mu0}
\ee
where we have included the matter coupling, which is indeed finite.  This is the exact Stuckelberg theory in this limit. Note that this action possesses a symmetry $\hat \chi \to \hat \chi +f(x^0)$, which is an artifact  of foliation preserving diffeomorphisms. The corresponding graviton theory goes like
\be
S_{graviton}=-\int d^4 \hat x ~\frac{1}{2}\hat h^{\mu\nu} {\cal E}\hat h_{\mu\nu}+\frac{l_{pl}}{\sqrt{c_s}}\int d^4\hat x ~\frac{1}{2}\hat h_{\mu\nu} T^{\mu\nu}
\ee
where $\hat h_{\mu\nu}=\frac{1}{\sqrt{c_s}}h_{\mu\nu}$. Again, the matter coupling is held finite. 

The Stuckelberg equations of motion are now given by
\be
2\del^2 \del_{\hat 0}^2 \hat \chi - 2\del^4 \hat \chi+ \frac{1}{\Lambda_{sc}^{2} } \del_i \left[ (\del_i \hat \chi ) \del^2 ( \del_j \hat \chi \del_j \hat \chi ) \right] - \frac{2}{\Lambda_{sc}} \del_i \left[  \del_i \del_j \del_{\hat 0} \hat \chi \del_j \hat \chi + \del_i \del_j \hat \chi \del_{\hat 0} \del_j \hat \chi + \del_{\hat 0} \del^2 \hat \chi \del_i \hat \chi  \right]=\frac{1}{\Lambda_{sc}} \del_\mu T^{\mu0} \label{chieq}
\ee
whereas the graviton equations of motion are given by
\be
{\cal E} \hat h^{\mu\nu}=\frac{l_{pl}}{2\sqrt{c_s}}T^{\mu\nu} \label{grav}
\ee
For our slowly varying point sources, we have
\be
T^{\mu\nu} \to M_*\delta^{(3)}(\vec x)\text{diag}(1, 0, 0,0),
\qquad
\del_\mu T^{\mu0} \to - \Gamma M_* \delta^{(3)}(\vec x)
\ee
with $\Gamma M_*$ and $l_{pl} M_*/\sqrt{c_s}$ held fixed. We shall seek static spherically symmetric solutions to equation (\ref{chieq}) of the form $\hat \chi=\hat \chi(r)$.  After integrating over a sphere of radius $r$, centred on the origin, we find that
\be
\frac{d}{dr} \left(\frac{1}{r^2} \frac{d}{dr} r^2 u\right)- \frac{1}{2 \Lambda_{sc}^{2} } \frac{u}{r^2} \left(\frac{1}{r^2} \frac{d}{dr}  u^2\right)=\frac{1}{ \Lambda_{sc} }\left( \frac{\Gamma M_\ast}{8 \pi r^2} \right)
\ee
 where $u={\hat \chi}'(r)$. We can solve this equation as a power series in $1/\Lambda_{sc}$.  To  all orders in the expansion, the unique solution is given by 
 $$
 u= - \frac{1}{16 \pi} \frac{\Gamma M_\ast}{\Lambda_{sc}}  \qquad \implies \qquad \hat \chi(r) = c- \frac{1}{16 \pi} \frac{\Gamma M_\ast}{\Lambda_{sc}} r
 $$
 where $c$ is some arbitrary integration constant. Owing to the foliation preserving diffeomorphisms,  the Stuckelberg mode possesses a shift symmetry $\hat \chi \to \hat \chi+const$. We use this symmetry to  set $c=0$, so that the final solution  is given by  
 \be
 \hat \chi(r) = - \frac{1}{16 \pi} \frac{\Gamma M_\ast}{\Lambda_{sc}} r
 \ee
Now consider the graviton equation (\ref{grav}). The solution to this equation is well known, and most conveniently expressed in Newtonian gauge,
\be
\hat h_{00}=\left(\frac{l_{pl} M_*}{\sqrt{c_s}}\right) \frac{1}{8\pi r}, \qquad \hat h_{ij}=\left(\frac{l_{pl} M_*}{\sqrt{c_s}}\right) \frac{1}{8\pi r} \delta_{ij}
\ee
Now suppose we probe the field generated by the source using a second point mass, with energy-momentum tensor satisfying
\be
\tilde T^{\mu\nu} \to \tilde M_* \delta^{(3)}(\vec x-\vec y)\text{diag}(1, 0, 0,0), \qquad \del_\mu \tilde T^{\mu0} \to - \tilde \Gamma \tilde M_* \delta^{(3)}(\vec x-\vec y)
\ee
with $\tilde \Gamma \tilde M_*$ and $l_{pl} \tilde M_*/\sqrt{c_s}$ held fixed in the relevant limits. The potential energy of the probe due to the Stuckelberg interaction is given by
\be
V_{\hat \chi}= \int \d^3   x ~\frac{\hat \chi}{\Lambda_{sc}} \del_\mu \tilde T^{\mu0}=\frac{1}{16 \pi} \frac{1}{\Lambda_{sc}^2} \Gamma M_\ast \tilde \Gamma \tilde M_\ast r
\ee
It follows that the Stuckelberg field mediates a constant attractive force 
\be
\vec F_{\hat\chi}=-\vec \nabla V_{\hat \chi}= -\frac{1}{16 \pi} \frac{1}{\Lambda_{sc}^2} (\Gamma M_\ast ) ( \tilde \Gamma \tilde M_\ast) \underline{\hat r} \label{Fchi}
\ee
 between the two point masses. Thus we have confinement, and in particular,  a bound universe. There is no fall off with distance, and no way for the masses to escape the mutual Stuckelberg force on each other. 

In contrast, the potential energy of the probe due to the graviton interaction is just the Newtonian potential
\be
V_{\hat h}=-\frac{l_{pl}}{\sqrt{c_s}}\int d^3 x ~\frac{1}{2}\hat h_{\mu\nu} \tilde T^{\mu\nu}=-\frac{l_{pl}^2}{16 \pi}\left(\frac{M_*\tilde M_*}{ c_s}\right)\frac{1}{ r}
\ee
with the usual attractive force satisfying an inverse square law, 
\be
\vec F_{\hat h}=-\vec \nabla V_{\hat h}=-\frac{l_{pl}^2}{16 \pi}\left( \frac{M_*\tilde M_*}{c_s}\right) \frac{\underline{\hat r}}{ r^2} \label{Fh}
\ee
Note that there is an additional factor of $c_s$ compared to the standard formula owing to the fact that we are using $\hat x^0=c_s x^0$ as our time coordinate. Now let us compare the two forces, given by  equations (\ref{Fchi}) and (\ref{Fh}). They both mediate an attractive force although they scale differently with distance.  At large distances  the Stuckelberg force dominates while at smaller distances the graviton force dominates.  The two forces are equal at a distance of
\be
r_{eq} =  \frac{ \Lambda_{sc} l_{pl}}{\sqrt{c_s\Gamma \tilde \Gamma}} = \frac{\lambda-1}{\sqrt{\Gamma \tilde \Gamma}}  . \label{req}
\ee
For $r \ll r_{eq}$ the Stuckelberg force is irrelevant, and one should expect to recover Newtonian gravity for a two particle system. However, we would like to stress that this is not really a Vainshtein effect since it is not the case that non-linear interactions screen the Stuckelberg force above a certain scale. In fact, non-linear interactions never become important in this particular example.  It is simply the case that the graviton force grows at short distances whereas the Stuckelberg force remains constant. 

Nonetheless, we  expect that provided we have a  large enough crossover scale for objects within the solar system, Newton's law always should always  hold at this scale. Indeed, if we assume that $\Gamma \sim \tilde \Gamma \sim H_0$, then  the crossover scale $r_{eq} \sim (\lambda-1)/H_0$. For $\lambda-1 \sim 10^{-10}$, we expect to recover Newton's law within the Oort cloud, at distances $r \ll 10^{16}$ m. One might worry that  that there are, in principle, many many far away sources for the Stuckelberg field  that   will exert a constant long range force on the objects within the solar system. However, as long as we assume homogeneity and isotropy at large scales, the effect of far away sources should cancel one another out. 

In summary then,  for point sources, with slowly varying mass, there is a scale at which the graviton force becomes dominant and one is able to recover Newtonian gravity. However,  this  is {\it not} a Vainshtein effect as non-linear interactions never play much of a role.  Does this mean that strong coupling is not important? Clearly this is unlikely to be the case in less symmetric configurations. It would be interesting to consider alternative sources, in particular a binary system that violates energy conservation. As was pointed out in \cite{us}, binary systems are particularly relevant as they represent a direct test of perturbative GR.
We conclude on a much more troubling note.  As is clear from equation (\ref{req}), the crossover scale generically depends on the variation rate of the probe,  $\tilde \Gamma$, as well as the rate of the source, $\Gamma$.  This illustrates the fact that the Stuckelberg force violates the Equivalence Principle. Indeed, probe masses with different  $\Gamma$'s will experience different accelerations in the presence of a Stuckelberg field generated by a point source. Such violations will be of the order $\eta \sim ( \Gamma_1-\Gamma_2)/(\Gamma_1+\Gamma_2)$ for different probes with violation rates $\Gamma_1, ~\Gamma_2$. The violation will kick in at large distances, beyond the lesser of the two crossover scales. We expect this to be a generic feature for objects that source the Stuckelberg field. Violation of Equivalence Principle can potentially be used to place phenomenological constraints on Ho\v rava gravity, a fact  that had not been noticed previously.

\section{Conclusions} \label{sec:conc}
Since its proposal just over a year ago, Ho\v rava's toy model of quantum gravity has attracted a huge amount of interest. By taking an appropriate decoupling limit we have obtained new insights into Ho\v rava gravity and its suitability as a quantum gravity candidate. Our analysis focuses on the  ``healthy extension'' of the theory~\cite{blas2}, and specifically the case where Lorentz violation occurs at the Planck scale. We have been particularly interested in the behaviour of the Stuckelberg field, the troublesome extra degree of freedom arising from breaking full diffeomorphism invariance. Taking the limit where this decouples from the gravity sector simplifies the calculations significantly. Indeed, both the validity and simplicity of this approach were clearly demonstrated  in section \ref{sec:ghosts}, where we recovered some known results~\cite{sotpap, blas2} with consummate ease. In particular, we reproduced the strong coupling result of \cite{sotpap} in an elegant manner. In some respects our analysis is more complete since we do not restrict attention to the low energy effective theory.  The correct way to interpret our result is to realise that the only way to avoid strong coupling is to explicitly introduce a new scale in the theory, below the would be strong coupling, giving rise to a much lower scale of Lorentz violation.

Indeed, it has already been proposed that strong coupling can be avoided if one accepts this slightly ad hoc introduction of a new scale in the theory~\cite{blas3}. Since this creates a hierarchy between the Lorentz breaking scale and the Planck scale, we might wish to avoid this on grounds of naturalness\footnote{See, however, footnote 9 of ref. \cite{blas3}}. Nevertheless  it would be interesting to consider some possible implications of this scenario, as we will discuss shortly. Returning to the case of Planckian Lorentz violation,  we find that the strong coupling scale is smaller than the scale of the gradient expansion, confirming the validity of the analysis in \cite{sotpap}. The largest interactions become strongly coupled at the scale, $\Lambda_{sc} \sim \frac{1}{l_\al} c_s^{3/2}$, which is less than the scale $1/l_\al$ corresponding to the cut-off in the derivative expansion.  Of course,  this argument relies on the fact that the Stuckelberg field does not propagate faster than light ($c_s \leq 1$). However, even if we allow superluminal propagation, which is not unreasonable in a Lorentz violating theory, strong coupling is still a problem. To see this note that  the cubic term $\sim \frac{1}{l_\al^2} \dot \chi (\del_i \dot \chi)^2$ becomes strongly coupled at scale $\frac{1}{l_\al} c_s^{-1/2}$, which is  below the higher derivative cut-off for $c_s > 1$. 

Why is strong coupling so bad? In principle, its not. It depends on the context. QED becomes strongly coupled in the UV due to the presence of a Landau pole, but the theory is still renormalisable. In contrast, we  have never known for sure if Ho\v rava gravity was a renormalisable theory. It was only ever suggested by a dubious power counting argument, in which one wrongly infers a schematic form  for the action in terms of  the perturbative degrees of freedom.  The problem is that the Stuckelberg mode is essentially ignored. Indeed, if we assume  that to leading order one can schematically replace the curvatures with derivatives of the graviton and that there is nothing else to worry about, one might expect the action to resemble those discussed in \cite{visser}, which {\it are} power counting renormalisable. But, of course, we should not ignore the Stuckelberg mode. Typically, the Stuckelberg theory behaves nothing like the renormalisable actions described in \cite{visser}. We therefore have  little reason to expect renormalisability and little reason to tout Ho\v rava gravity as a UV complete theory of gravity. To get the Stuckelberg theory to behave more appropriately in the UV, we need to introduce a new scale of Lorentz violation by hand, and take it to be well below the Planck scale, as suggested by \cite{blas3}. 

Of course, even if it is not UV complete, one could ask if Ho\v rava gravity is a phenomenologically viable modification of GR. What is the significance of strong coupling in this context? The  strong coupling scale is the scale at which quantum fluctuations on the vacuum start to interact strongly. In the presence of a perturbative source, this scale can be linked to the scale at which classical linearised perturbation theory breaks down\footnote{In DGP gravity, the strong coupling scale is given by $(M_{pl}H_0^2)^{1/3}$~\cite{dgpsc}. This scale can be linked to the scale at which classical perturbation theory breaks down around a  heavy  source. Indeed, for a source of mass $M$, linearised perturbation theory breaks down at a scale $(M H_0^2)^{(1/3)}$\cite{dgpvain}. }. In \cite{us}, we argued that the original Ho\v rava theory could not be viable since it was strongly coupled on all scales. This  was a problem because it meant that there was no scale at which one could apply the standard linearised theory around a heavy source. Linearised General Relativity around a heavy source has been well tested, at least indirectly, thanks to the Nobel Prize winning binary pulsar observations of Hulse and Taylor~\cite{ht}. In the extended version of Ho\v rava gravity with Lorentz violation at Planckian scales,  strong coupling kicks at a finite scale $\Lambda_{sc} \sim \frac{1}{l_\al} c_s^{3/2}= \frac{1}{l_{pl}}\left(\frac{(\lambda-1)^3}{\al}\right)^{1/4} $.  This suggests that linearised theory around a heavy source is valid up to some finite scale, although one clearly ought to check that the Stuckelberg field does not spoil GR's successful matching to Hulse and Taylor's observations (see \cite{will} for corresponding studies in Brans-Dicke gravity).

However, these arguments are not quite enough to rule out strongly coupled versions of Ho\v rava gravity on phenomenological grounds since they ignore  any possible Vainshtein effect~\cite{vain}. The Vainshtein effect typically occurs in modifications of GR that exhibit strong coupling. Even if one has too many degrees of freedom at the linear level to mimic General Relativity, non-linear interactions can save the day. Because of strong coupling, bound states form, allowing extra degrees of freedom to be screened and enabling one to recover GR at short enough distances. To study any possible Vainshtein mechanism in Ho\v rava gravity, it is important to understand how the Stuckelberg mode couples to matter. We have been able to determine this  coupling by making use of the reduced diffeomorphisms, as described at the end of section \ref{sec:gen}. It turns out that one should include some violation of energy-momentum conservation, measured by some scale $\Gamma$, to source the Stuckelberg field. Note that this is not as crazy as it might sound. Energy-momentum conservation is not required because we don't have full diffeomorphism invariance. From a phenomenological perspective, we can assume that  $\Gamma \lesssim H_0$, so that violations only become apparent on superhorizon scales.

In section \ref{sec:vain} we studied the interaction between two point particles, with slowly varying masses. This is probably the simplest way to source the Stuckelberg field, in order to see if  there is indeed any sort of Vainshtein mechanism at work.  Using our results from strong coupling, and taking the limit $c_s \to 0$ while keeping $\Lambda_{sc}$ finite, we were able to write down the exact action for this system.  It turns our that the Stuckelberg field gives rise to a constant attractive force between the two particles. This field dominates over the graviton force at large distances and gives rise to confinement.  At short distances the graviton force dominates and one can recover Newtonian gravity. However this is not a Vainshtein effect since non-linearities do not play any role in screening the Stuckelberg force. In fact, non-linearities play no role at all in this example, although we do not expect this to be true in general. In fact, it is probably just an artifact of our taking the (almost) static limit. It would be interesting to consider alternative sources for the Stuckelberg field, most notably a binary system that weakly violates energy-momentum conservation.

In the absence of a Vainshtein mechanism in our example, perhaps the most important result of section \ref{sec:vain} was the realisation that Ho\v rava gravity will inevitably lead to violations of the Equivalence Principle. This is because the Stuckelberg force depends  on the rate of energy-momentum conservation violation $\Gamma$ of each particle, as well as their masses. Different probes with different $\Gamma$'s will therefore feel different accelerations. This gives a non-trivial E\"otv\"os parameter, for which we have very tight experimental bounds~\cite{eotvos}.   Of course, one might hope to evade this issue by imposing, without adequate motivation, that only conserved sources are allowed in this theory. Whilst this can be achieved through a specific choice of matter coupling in the classical Lagrangian, it is clear that loop effects will introduce small corrections, suppressed by the scale of Lorentz symmetry breaking. Even though the resulting ``scale of non-conservation", $\Gamma$ is small for a generic source, it will crucially be non-zero. This can lead to large effects since the violations of Equivalence Principle will be of the order $\eta \sim ( \Gamma_1-\Gamma_2)/(\Gamma_1+\Gamma_2)$  for probes that  violate energy-momentum  conservation at different rates  $\Gamma_1 \neq \Gamma_2$. Of course, the effect only kicks in beyond the lesser of the two crossover scales (\ref{req}) for each probe, so this phenomena could be used to place a lower bound on the value of $\lambda$. 

Although some of our results hold only in the ``healthy extension'' of the Ho\v rava gravity~\cite{blas2} {\it with  Planckian Lorentz violation}, it is clear that all the analysis could be repeated fairly simply for the original non-projectable theory, with roughly similar conclusions.  One might even consider extending our method to  the projectable theory, by adding a term $S_{gf}=\int dt d^3x \sqrt{\gamma}N Q_i a^i$ into the original action. $Q_i$ is a Lagrange multiplier, whose equations of motion enforce $a_i = 0 \Rightarrow D_i \log N = 0 \Rightarrow N = N(t)$. Indeed, this might  well be a worthwhile exercise given the recent claims that this version of the theory is free from pathologies when one considers fluctuations about de Sitter as opposed to Minkowski~\cite{projok}.

Let us now turn to the issue of scales. Note that we have avoided the introduction of additional scales into the theory on ground of  naturalness. This manifests itself through the absence of large dimensionless coefficients in the action, and a single Lorentz violating scale $l_{UV}$, which is taken to be Planckian. In \cite{blas3}, it is claimed that the strong coupling problems discussed here can be avoided by introducing a dimensionless coefficient $B \gtrsim 10^7$. This means that  Lorentz invariance is broken at much lower scale $M_* \ll M_{pl}$ than one might expect. We could modify our decoupling limit by holding $M_*$ fixed as we take $l_{pl} \to 0$, but this would still not give the desired dispersion relation for the Stuckelberg mode in the UV ($w^2 \propto k^6$) required to ``cure" strong coupling. Our conclusion then is that this effect can not be reproduced in the decoupling limit -- one needs to retain some coupling between the graviton and the Stuckelberg mode, even if it is just to quadratic order.  What we can do is ask what impact this low scale of Lorentz violation has on tests of Equivalence Principle.  As we have discussed, even if one assumes conserved sources classically, quantum mechanically we will get violations of energy-momentum conservation, suppressed by some power of the Lorentz symmetry breaking scale. The lower the scale, the less the suppression, and the larger the generic value of $\Gamma$. This would raise the lower bound on $\lambda$ derived from the crossover scale. It would be interesting to see if this can be made compatible with tests of Lorentz violation which place an upper bound on the value of $\lambda$ (see for example, \cite{blas3}).

In conclusion, making use of the powerful tools of the decoupling limit has allowed us to demonstrate that one cannot avoid strong coupling problems in the ``healthy" extension of Ho\v rava gravity {\it if one assumes Lorentz violation at the Planck scale}.  With additional pressure coming from experimental observation, one is forced to introduce a much lower  scale of Lorentz violation, along the lines proposed by \cite{blas3}. Perhaps surprisingly, the details of avoiding strong coupling in that scenario cannot be captured by the decoupling limit.    At the level of  phenomenology, we have studied the force between two point particles with slowly varying masses. We have found no Vainshtein effect, but we have seen violations of the Equivalence Principle. We believe the latter is a generic feature, but not the former.  It is possible that tests of Equivalence Principle will  present a challenge to the low scale of Lorentz violation designed to cure strong coupling~\cite{blas3}, although a more detailed study is clearly required.  

\subsection*{Acknowledgements}
We would like to thank Ed Copeland, Paul Saffin, Gustavo Niz, Nemanja Kaloper, Kazuya Koyama, Thomas Sotiriou and Antonios Papazoglou
for useful discussions. AP is funded by a Royal Society University
Research Fellowship and IK by an STFC studentship. 
 
\appendix

\section{Derivatives of potential terms}\label{sec:potdevs}
Below are the first derivatives of the potential $\vfour$.
\bea
\partd{\vfour}{a_{\mu}} &=& 4A_1 ( a^\nu a_\nu) a^{\mu} + 2 A_2 a^\mu a^\nu{}_\nu\\ &&+ 2B_1( \K_{\mu\nu} \K^{\mu\nu} - \K^2) a^\mu + 2 B_2( \K_{\mu \alpha} \K^\al_\nu- \K \K_{\mu\nu}) a^\nu \nonumber \\
\partd{\vfour}{a_{\mu \nu}} &=& A_2 ( a^\rho a_\rho) \eta^{\mu \nu} + 2 A_3 ( a^\rho{}_\rho) \eta^{\mu \nu} + 2 A_4 a^{\mu\nu} \nonumber\\ 
&& +B_3( \K_{\sigma \rho} \K^{\sigma \rho} - \K^2) \eta^{\mu \nu} \\
\partd{\vfour}{\K_{\mu \nu}} &=& 2 B_1( \K^{\mu\nu} - \K \eta^{\mu \nu}) a^\rho a_\rho + B_2( 2 \K^{(\mu}_{\rho} \delta^{\nu)}_\sigma - \eta^{\mu\nu} \K_{\rho \sigma} - \K \delta^{(\mu}_{\rho} \delta^{\nu)}_{\sigma}) a^\rho a^\sigma \nonumber \\
&& + 2 B_3( \K^{\mu\nu} - \K \eta^{\mu \nu} ) a^\rho{}_\rho + 4 C_1( \K_{\rho \sigma} \K^{\rho \sigma} - \K^2) ( \K^{\mu\nu} - \K \eta^{\mu \nu} )
\nonumber \\
&&
+2 C_2 ( 2 \K_{\al}^{(\mu} \delta^{\nu)}_\rho - \eta^{\mu \nu} \K^{\al}_{\rho} - \K \delta^{(\mu}_{\rho} \eta^{\nu) \al} ) ( \K^{\rho}_{\beta} \K^{\beta}_{\al} - \K \K^{\rho}_{\al} )
\eea

\section{Determining the strong coupling scale}\label{sec:lowestscale}

In section \ref{sec:fluc}, we stated that the appropriate strong coupling scale was given by the interaction term in equation (\ref{s4}), leading to a strong coupling scale of $\Lambda_{sc} \sim \frac{1}{l_\al}c_s^{3/2}$. Here we will demonstrate that this is the appropriate scale, by virtue of having the highest power of $c_s$. Recall that the scales at which the various terms become strongly coupled are given by equation (\ref{strongcouplingscales}), which we repeat here,
\be \tag{\ref{strongcouplingscales}}
\Lambda^\al_{(a, n)}=\frac{1}{l_\al}c_s^{\frac{1}{2}+\frac{2-a}{n-2}}, \qquad
\Lambda^\K_{(a, n)}=\frac{1}{l_\al}c_s^{\frac{1}{2}-\frac{a}{n-2}}, \qquad
\Lambda^V_{(a, n)}=\frac{1}{l_\al}c_s^{\frac{1}{2}+\frac{1-a}{n}}.
\ee
Let us begin by considering the $\Lambda^\al_{(a, n)}$ terms,
\begin{figure}[htb]
\includegraphics[width=0.7\textwidth]{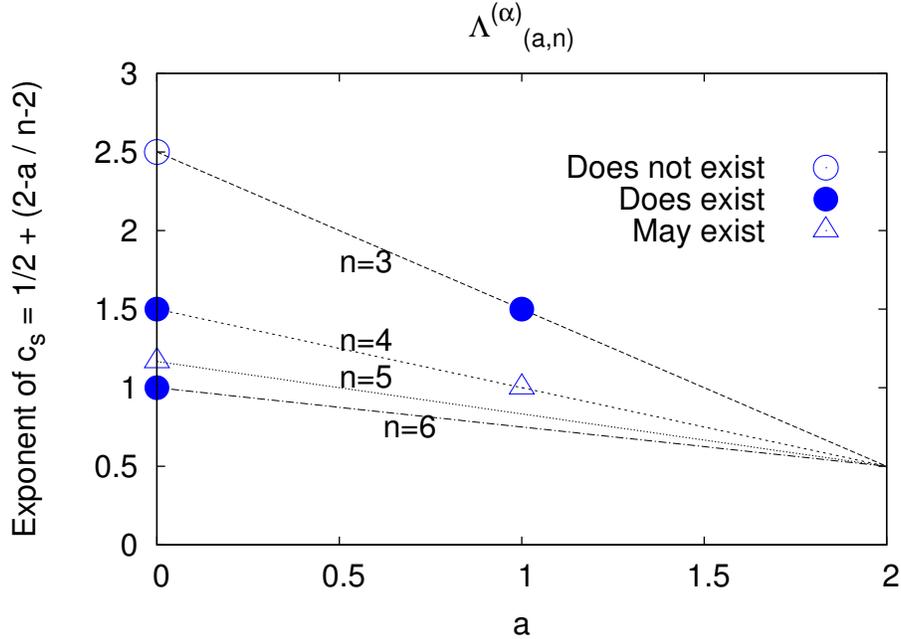}
\caption{The exponent of $c_s$ to which various potential strong coupling terms in $\Lambda^\al_{(a, n)}$ are raised.}
\label{fig:strongc}
\end{figure}
We want to identify the terms with the highest powers of $c_s$, which will result in the lowest strong coupling scale, since $c_s \leq 1$. It is also necessary to bear in mind that $n$, $a$ are restricted to be integers and that not all the `possible' terms appear in the perturbative expansion. The exponent of $c_s$ in the $\Lambda^\al_{(a, n)}$ terms is plotted in Figure \ref{fig:strongc}. It is clear that the strongest coupling would be given by the term $(a=0,n=3)$, but this is not present in the expansion. Instead, the two terms corresponding to $(a=1,n=3)$ and $(a=0,n=4)$ have the greatest exponent of $c_s$ of any terms present and so result in the lowest energy scale $\Lambda^\al_{(1, 3)} = \Lambda^\al_{(0, 4)} = \Lambda_{sc} \sim \frac{1}{l_\al}c_s^{3/2}$. Furthermore, it is clear from equation (\ref{strongcouplingscales}) that that is the smallest scale, since $\min \Lambda^\K_{(a, n)} = \frac{1}{l_\al} c_s^{1/2} = \Lambda_{sc}/c_s$ and $\min \Lambda^V_{(a, n)} = \frac{1}{l_\al} c_s^{5/6} = \Lambda_{sc}/c_s^{4/6}$ will always be larger than $\Lambda_{sc} $ for $c_s \leq 1$.

Thus the terms which lead to strong coupling, and set the strong coupling scale, are the $(a=1,n=3)$ and $(a=0,n=4)$ terms from $S^\al_{(a,n)}$.

\end{document}